\documentclass[10pt,sigconf,letterpaper,nonacm,natbib=false]{acmart}

\usepackage{subcaption}

\usepackage{hhline}
\usepackage{graphicx}
\usepackage{textcomp}
\usepackage{esvect}
\usepackage{xcolor}

\def\BibTeX{{\rm B\kern-.05em{\sc i\kern-.025em b}\kern-.08em
    T\kern-.1667em\lower.7ex\hbox{E}\kern-.125emX}}
\usepackage{algorithm}
\usepackage{algcompatible}
\usepackage{booktabs}
\usepackage{diagbox}    
\usepackage{array}
\usepackage{lipsum}
\usepackage{mathtools}

\newcolumntype{L}[1]{>{\raggedright\let\newline\\\arraybackslash\hspace{0pt}}m{#1}}
\newcolumntype{C}[1]{>{\centering\let\newline\\\arraybackslash\hspace{0pt}}m{#1}}
\newcolumntype{R}[1]{>{\raggedleft\let\newline\\\arraybackslash\hspace{0pt}}m{#1}}
\AtBeginDocument{%
  \providecommand\BibTeX{{%
    Bib\TeX}}}

\setcopyright{acmcopyright}
\copyrightyear{2018}
\acmYear{2018}
\acmDOI{XXXXXXX.XXXXXXX}

\RequirePackage[
  datamodel=acmdatamodel,
  style=acmnumeric,
  ]{biblatex}

\begin{document}

\title{Predictive Edge Caching through Deep Mining of Sequential Patterns in User Content Retrievals}

\author{Chen Li}
\email{chen.lee@nyu.edu}
\affiliation{%
  \institution{New York University}
  \country{USA}
  \postcode{11201}
}

\author{Xiaoyu Wang}
\email{xw2597@nyu.edu}
\affiliation{%
  \institution{New York University}
  \country{USA}
}

\author{Tongyu Zong}
\email{tz1178@nyu.edu}
\affiliation{%
  \institution{New York University}
  \country{USA}
}

\author{Houwei Cao}
\email{hcao02@nyit.edu}
\affiliation{%
  \institution{New York Institute of Technology}
  \country{USA}
}

\author{Yong Liu}
\email{yongliu@nyu.edu}
\affiliation{%
  \institution{New York University}
  \country{USA}
}





\begin{abstract}
Edge caching plays an increasingly important role in boosting user content retrieval performance while  reducing redundant network traffic.
The effectiveness of caching  ultimately hinges on the accuracy of predicting content popularity in the near future. 
However, at the network edge, content popularity can be extremely dynamic due to diverse user content retrieval behaviors and the low-degree of user multiplexing. It's challenging for the traditional reactive caching systems to keep up with the dynamic content popularity patterns. 
In this paper, we propose a novel Predictive Edge Caching (PEC) system that predicts the future content popularity using  fine-grained learning models that mine sequential patterns in user content retrieval behaviors, and opportunistically prefetches contents predicted to be popular in the near future using idle network bandwidth. 
Through extensive experiments driven by real content retrieval traces, we demonstrate that PEC can adapt to highly dynamic content popularity at network edge, and significantly improve cache hit ratio and reduce user content retrieval latency over the state-of-art caching policies. 
More broadly, our study demonstrates that edge caching performance can be boosted by deep mining of user content retrieval behaviors.
\end{abstract}


\maketitle

\section{Introduction}
Emerging applications, such as Virtual/Augmented/Mixed Reality, require high-throughput and low-latency content delivery. Edge caching is a promising solution to simultaneously reduce user content retrieval latency and mitigate traffic congestion in core networks. The key to 
achieve high caching gain is to accurately predict the content popularity in the near future. The classic caching policies, such as LFU, 
LRU and their variants, assume contents that are popular in the past will continue to be popular in the near future. Caching replacement is therefore guided by simple statistics of the past content requests, such as time elapsed since the last request (LRU) and the frequency of past requests (LFU). Compared with the traditional CDN servers, each edge cache node is equipped with smaller storage and serves a smaller user group. As a result, the aggregate content popularity of users served by an edge cache node is less stationary, and more difficult to be accurately estimated by simple aggregate statistics of the past requests. 

Lots of efforts have been made recently to address the edge caching challenge. Some methods, e.g. CRFP~\cite{CRFP} and SG-LRU~\cite{SGLRU}, have been proposed to improve or combine LFU and LRU policies.  Recently, machine learning methods have been applied to improve caching performance by explicitly or implicitly learning the future content popularity, e.g. LHR\cite{yan2021learning}, Learning Relaxed Belady (LRB)~\cite{song2020learning}, and  CEC~\cite{CEC_tongyu}. But all of these methods are still  reactive caching, in which cache replacements are only triggered by cache misses. Meanwhile, proactive caching, e.g.  \cite{proactive5Grole}, enjoys the freedom of prefetching any content at any time with additional bandwidth cost. As discussed in \cite{proactiveneedpopulairty}, content popularity estimation is critical for efficient proactive caching. \cite{optimalproactive} gives the theoretically  upper bound for proactive caching when content popularity is stationary. Periodical proactive cache updates, e.g. \cite{predicalproactive}, can cope with non-stationary content popularity. However, it cannot adapt to the content popularity variations between two updates. 

\begin{figure}[htb]
    \centering
    \includegraphics[scale=0.2]{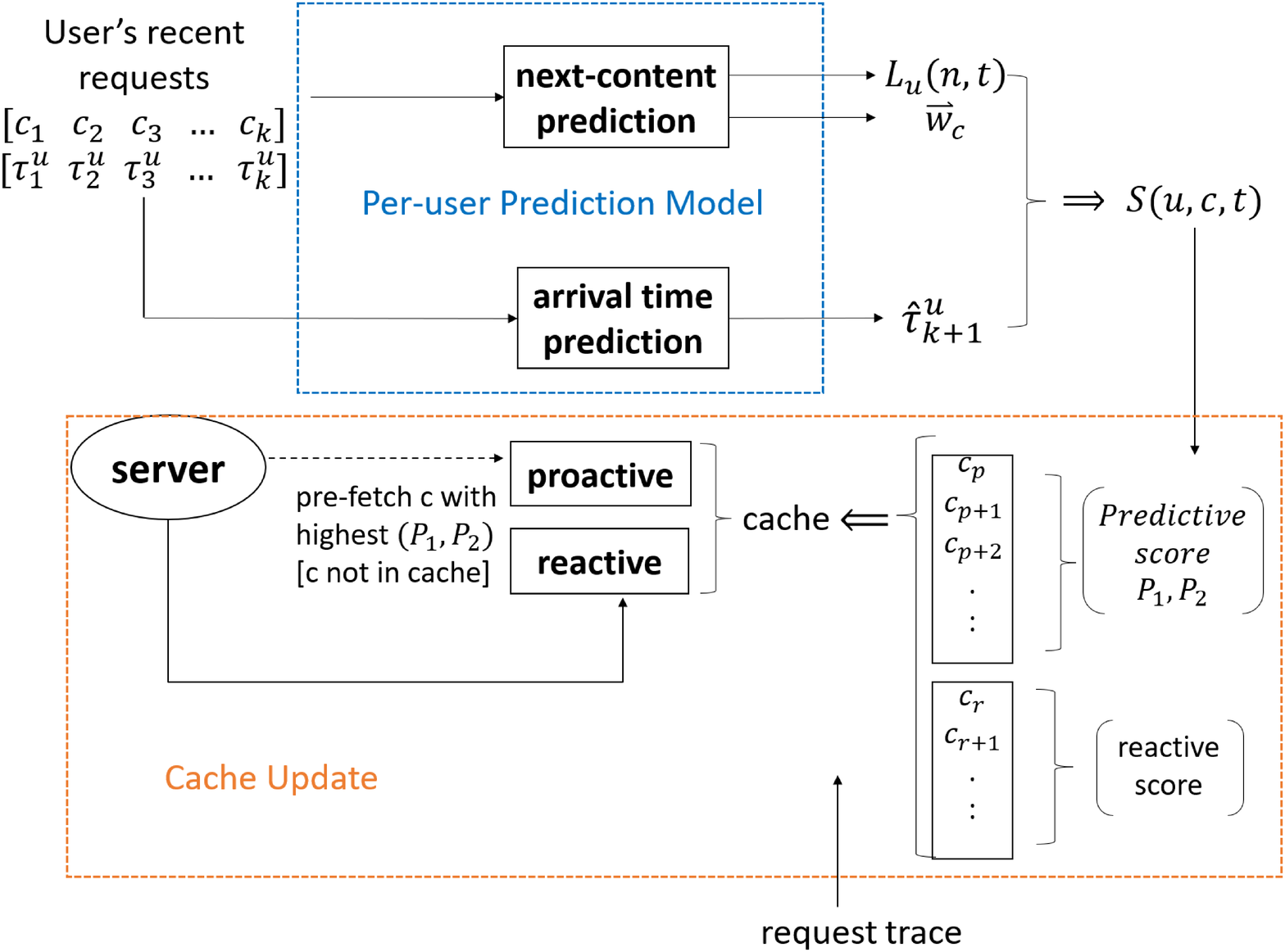}
    \caption {Overview of PEC System. The upper part is per-user next-content prediction model. Given user's recent requests with timestamps, the predicted contents $L_u(n,t)$ and their weights $\protect\vv w_c$ can be obtained. With the estimated arrival time $\hat{\tau}^u_{k+1}$, a real time predictive score $S(u,c,t)$ will be generated. The lower part is realtime caching policy with caching scores. The cache is partitioned into proactive portion and reactive portion, which are updated according to the predictive and reactive caching scores respectively.}
    \label{fig:system}
\end{figure}

{\it In this paper, we propose a novel Predictive Edge Caching (PEC) system that predicts the future content popularity using fine-grained learning models that mine sequential patterns in user content retrieval behaviors, and opportunistically pre-fetches contents predicted to be popular in the near future to improve cache hit ratio and reduce content retrieval latency.} To address the diverse user content interests and content consumption behaviors, instead of using the aggregate content request statistics of all users, we mine each individual user's content request history and predict when and which content each user is likely to request the next using sequential machine learning models. We then aggregate the next-content predictions of all users to generate predictive caching scores reflecting the future content popularity. Contents predicted to be popular will be proactively prefetched into cache in the background using spare network bandwidth. To closely keep track of  dynamic content popularity, per-user next-content predictions, predictive caching score updates, and proactive content prefetches are all conducted in realtime. The high-level structure of PEC is illustrated in Figure~\ref{fig:system}.  Within this novel predictive edge caching framework, we make the following contributions:  
\begin{itemize}
    \item We develop machine learning models to mine the user sequential viewing patterns. We show that n-gram and self-attention sequential models have complementary performance, and can be easily combined to generate accurate fusion prediction. We use simple, yet robust, statistical methods to predict the arrival time of the next request.  
   \item To guide online prefetching, we develop models to aggregate per-user content request predictions into per-content predictive caching scores, and update them continuously as posterior probabilities over time.  
 \item We design a hybrid caching system that prefetches contents with high predictive scores into the proactive portion and caches popular contents missed by the predictions into the reactive portion. The sizes of the two portions are dynamically adjusted, and the content replacements in the two portions are orchestrated to maximize the caching gain. 
 
 \item  We develop a three-level strategy that controls the bandwidth consumption of proactive downloading 
 to minimize its negative impact on the regular traffic.
 
 \item Through extensive edge caching simulations driven by real-world data  traces, we demonstrate that, compared with the state-of-art reactive and the  traditional periodical proactive caching policies, PEC can significantly improve the cache hit ratio and reduce user content retrieval latency with controlled bandwidth overhead.   
\end{itemize}

The rest of the paper is organized as the following. In Section~\ref{Related Work}, we introduce the related work. 
 Sequential models for the next-content prediction are developed in Section~\ref{What to Cache: Sequential Prediction Models}, and Section~\ref{When to Cache: Estimate Arrival Time} presents the statistical model for the next request arrival time prediction. Section~\ref{sec:Caching} presents the design for using per-user predictions in hybrid caching. We perform comprehensive evaluation using real world data in Section~\ref{Evaluation}. And the final Section~\ref{Conclusion} delivers the conclusion.
 \vspace{-0.1in}
\section{Related Work}
\label{Related Work}
To meet the new challenges of content delivery, more and more researchers are focusing on edge caching in different ways. Caching methods can be roughly classified into two types: reactive caching and proactive caching. Reactive caching approaches, such as LRU, LFU, First-In First-Out (FIFO) and Greedy-Dual-Size-Frequency (GDSF)~\cite{fofack2012analysis, melazzi2014general,chan2000fifo, cherkasova1998improving} , replace the cached content having the lowest score with the requested content upon each cache miss, while proactive caching approaches update the whole cache periodically~\cite{dehghan2015complexity,poularakis2016exploiting,yang2015analysis,gregori2016wireless}. In \cite{lookahead}, authors investigated LRU, LFU and Belady's algorithms, and concluded caching for video streaming can benefit from look-ahead technique. AViC in \cite{avicpredictrequesttime} estimates chunk request time and evicts the furthest chunk when updating cache. In \cite{prefetchingnetworklevel}, contents are prefetched to edge cache nodes based on the aggregated content consumption statistics, instead of per-user content prediction.  Our work develops an adaptive real-time proactive caching approach which prefetches contents to become popular using idle bandwidth to adapt to dynamic popularity at edge.

Machine learning (ML), as a modern powerful tool, has also been well introduced into the caching field. Some efforts have been made on estimating dynamic content popularity~\cite{zhang2019toward,wei2021wireless,abolhassani2021single,song2020learning, li2018data, yan2021learning}, while other efforts have applied deep reinforcement learning (DRL) to directly generate content placement strategy given various evolving state features~\cite{wang2020federated,somuyiwa2018reinforcement,zhong2018deep,kirilin2020rl,fan2021pa,sadeghi2019reinforcement,sengupta2014learning,jiang2019multi,sadeghi2017optimal,zhu2018deep,wu2019dynamic,wang2020intelligent,zhong2020deep,ye2021joint,sadeghi2019deep, CEC_tongyu}.  In~\cite{predicalproactive}, LEAP , a machine learning model is trained to prefetch video segments to improve user's QoE in adaptive video streaming. Most of the proposed machine learning models are trained to learn the aggregated content consumption patterns of a group of users served by the same cache.
LRB~\cite{LRB_zhen} is a machine learning model to estimate the arrival time of the next request for a content (from any user in the group) within or out of the Belady boundary. Our machine models are designed to mine the sequential patterns of how each individual user consumes contents and predict for each user which content she will consume the next.  Nearly Optimal Cache (NOC) in ~\cite{NOC_zhou} aims to minimize the dynamic regret, which is the performance gap between an online learning algorithm and the best dynamic policy in hindsight. NOC has provably good worst-case performance for dynamic environments with no prior distribution assumptions, but it potentially degrades the performance when working with friendly request patterns.
Sequential prediction model is a hot topic in both industry and academia, ranging from the traditional Markov chain model~\cite{he2016fusing}, to the recent machine learning models, such as recurrent neural network (RNN)~\cite{hidasi2018recurrent}, long short-term memory (LSTM)~\cite{wang2020deep}, convolutional neural network (CNN)~\cite{wang2019towards}, and transformer~\cite{attentionisallyouneed} with self-attention mechanism. Self-attention model~\cite{kang2018self} outperforms some state-of-art sequential prediction models. In \cite{Tisasrec}, temporal-aware self-attention model delivers a promising prediction accuracy.\\

\section{Per-user Next Content Request Prediction}
The key for achieving high caching gain is to accurately predict which contents will be popular in the near future. We estimate the short-term content popularity on an edge cache node by predicting the next content that will be requested by each user served by the edge cache. Given the past content requests generated by user $u$, our model predicts {\it 1) which content user $u$ will request next 2) when the next request will be generated.}  More specifically, given the content request history of user $u$, 
\[\mathcal R^{(u)}(t) \triangleq \{ \langle \tau_1^{(u)}, c_1^{(u)}\rangle \cdots  \langle \tau_i^{(u)}, c_i^{(u)}\rangle, \cdots  \langle \tau_k^{(u)}, c_k^{(u)}\rangle\},\]
where $\tau_i^{(u)}$ and $c_i^{(u)}$ are the arrival time and content of the $i$-th request respectively, we want to predict $\tau_{k+1}^{(u)}$ and $c_{k+1}^{(u)}$ for the next request from user $u$.

\subsection{Sequential Models for Next-content Prediction}
\label{What to Cache: Sequential Prediction Models}
How a user sequentially consumes contents is highly dependent on the type of contents. For example, after $u$ finishes episode $m$ of 
a TV series $A$, there is a good chance that $u$ will move on to the $(m+1)$-th episode of $A$ the next. We can use a simple heuristic model to predict the next content for users watching TV series:    
\[P(c_{k+1}^{(u)}=A_{m+1}| c_{k}^{(u)}=A_{m}) =1,\]
where $A_m$ denotes episode $m$ of TV series $A$. We applied the simple heuristic to our datasets~\footnote{The datasets will be described in detail in Section~\ref{dataset}}, the prediction accuracy is 45.31\%.
%
%
%
%
However, there is no such strong sequential patterns for the other types of contents, such as movies, shows, short videos, etc. We now develop learning-based sequential models for the next-content prediction. 
\subsubsection{n-gram model}
\begin{figure}[htb]
    \centering
    \includegraphics[scale=0.35]{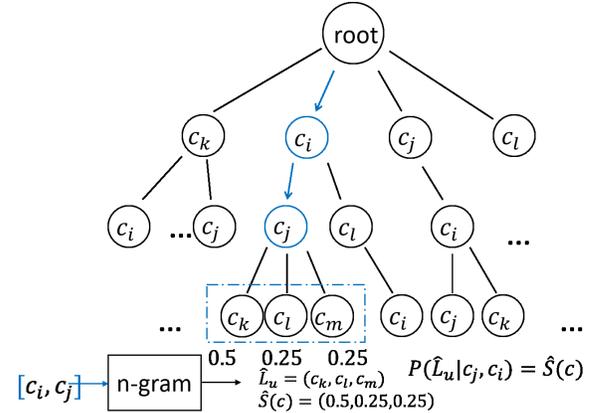}
    \caption{3-gram Model: after building the 3-gram tree, given input sequence $[c_i, c_j]$, model predicts the next-content candidates $\hat{L}_u$ with their probabilities.}
    \label{fig:ngramstructure}
\end{figure}

Sequential models are widely used in Natural Language Processing. We adopt the simple, yet powerful, n-gram model to solve our problem. More specifically, by assuming the next content only depends on the previous $n-1$ contents, the probability of the 
next content can be estimated by the conditional probability of  
\begin{equation}
\label{eq:ngram}
P(c_{k+1}^{(u)}=c | c_{k}^{(u)}, c_{k-1}^{(u)},  \cdots, c_{k-n+2}^{(u)}).
\end{equation}
Empirical conditional probability is derived from the history data. An illustration of 3-gram is shown in Fig.\ref{fig:ngramstructure}.

%
%
%

\subsubsection{Time-aware Self-attention Sequential (TSAS) Model}
The n-gram Markovian model can capture short-range content transition patterns, but falls short to model complex transition patterns in longer ranges. Meanwhile, Deep neural networks, such as Recurrent Neural Networks (RNNs) and Convolutional Neural Networks (CNNs), can be used to mine long-range patterns, but they perform well only with dense data. More recently, self-attention based sequence-to-sequence models, such as Transformer~\cite{attentionisallyouneed}, have achieved state-of-the-art performance in various NLP tasks. The main idea is to learn the ``self-attention weights" that quantify the pairwise impacts of words in the same sentence and predict the future words based on the previous words. The self-attention mechanism has been extended for sequential recommendation by additionally incorporating positional information and time information~\cite{Tisasrec}~\cite{kang2018self} in user-item interaction sequence. 

Caching is highly time-sensitive. We not only need to place popular contents in the cache, but also should do it at the `right' time. 
Contents to become popular the next day don't have to be cached now. Meanwhile, other than the watched contents, the time a user spent on each content also tells a lot about the user's content preference and watching habit.  For example, if a user often skips to another video within 10 minutes, it makes more sense to predict she will watch a short video instead of a long movie next. As a result, to accurately predict the next content $c_{k+1}^{(u)}$, we should not only consider the user's past content sequence, $\{ c_1^{(u)}, \cdots, c_{k}^{(u)}\}$, but also the timestamps of those contents $\{\tau_1^{(u)}, \cdots, \tau_{k}^{(u)}\}$. 


Motivated by the work in~\cite{Tisasrec}, we develop a customized time-aware self-attention model for next-content prediction. We convert each user's content request history into sequences of length $n$. For the clarity of presentation, we now denote one sequence from any user as $\{\langle \tau_1, c_1 \rangle \cdots \langle \tau_n, c_n \rangle\}$. We train a self-attention model using length-n sequences from all users\footnote{Padding will be applied if a user has requested less than $n$ contents.}. We assume that the impact of the $j$-th request of the sequence on the $i$-th request (We only consider $j < i$ for the causality consideration) depends on: 1) the requested content $c_j$,  2) $\tau_i-\tau_j$, the time elapsed from the $j$-th request to the $i$-th request. 

\begin{figure}[htb]
    \centering
    \includegraphics[scale=0.22]{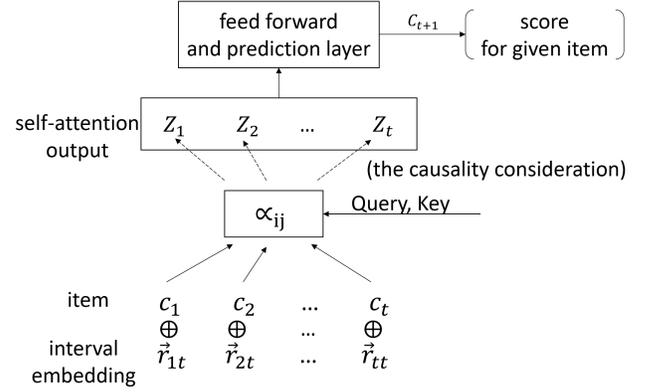}
    \caption{TSAS Model  Structure.{\bf{}}}
    \label{fig:TSASstructure}
\end{figure}

As illustrated in Figure~\ref{fig:TSASstructure}, to learn the self-attention weights, we embed contents using an embedding matrix ${E}^{(c)} \in {\mathbb{R}}^{|C| \times d}$, where \textit{d} is the dimension of the latent embedding space. A content embedding vector is projected to the corresponding \texttt{Key}, \texttt{Query} and  \texttt{Value} vectors using learnable projection matrices $W^k, W^q, W^v 
\in {\mathbb{R}}^{d \times d}$ respectively. 
All the time intervals are quantized and capped to be integers within $[0,k]$, and then embedded using two matrices ${E}^{(t,k)}, {E}^{(t,v)} \in {\mathbb{R}}^{k \times d}$, one for \texttt{Key}, the other for \texttt{Value}. We summarize the context of the first $i-1$ requests as a weighted sum of the embedding value vectors of the requested contents and the time-intervals between requests:  
\begin{equation}
    z_i = \sum^{i-1}_{j=1} \alpha_{ij}(E^{(c)}_{c_j} W^v + {E}^{(t,v)}_{\tau_{i}-\tau_j}),
\end{equation}
where $E^{(*)}_l$ represents the $l$-th row vector of the embedding matrix $E^{(*)}$, 
and the self-attention weight coefficient ${\alpha }_{ij}$ between request  \textit{i} and \textit{j} is calculated by the softmax function:

\begin{align*}
	{\alpha }_{ij} &= \frac{\exp {({m}_{ij}})}{\sum\nolimits_{k=1}^{n}\exp {({m}_{ik}})} \\
	{m}_{ij} &= \frac{E^{(c)}_{c_i} W^q {{( E^{(c)}_{c_j} W^k +{E}^{(t,k)}_{\tau_{i}-\tau_j})}^{T}}}{\sqrt{d}}. 
\end{align*}
We add non-linearity by feeding the output of the self-attention layer to a point-wise Feed-Forward Network (FFN) with drop-out and layer normalization. Finally, the probability of content $c$ for the $i$-th request is predicted as:
\begin{equation}
\label{eq:TSAS}
    P(c_i=c | \langle \tau_1, c_1 \rangle \cdots \langle \tau_{i-1}, c_{i-1} \rangle) \sim  \mathbf Z_i \cdot E^{(c)}_{c}, 
\end{equation}
where $\mathbf Z_i$ is the latent context vector outputted by the self-attention layer and FFN, and $E^{(c)}_{c}$ is the content embedding vector for candidate $c$.

\begin{figure}[htb]
    \centering
    \includegraphics[scale=0.4]{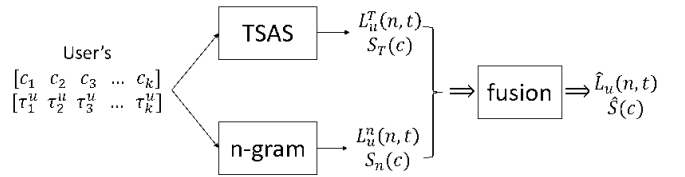}
    \caption{Fusion Prediction System.{\bf{}}}
    \label{fig:prediction}
\end{figure}




\subsubsection{Fusion Prediction Model}
\label{sec:fusion}
 To combine the short-range and long-range sequential patterns captured by the n-gram and TSAS models, we generate final prediction using model fusion. We tried different fusion methods, including CombSum, combMNZ and Reciprocal Rank Fusion(RRF)~\cite{liang2014fusion}. Among them, combSum fusion achieved the best performance and it has ranking scores for the final output, which can be used as content caching scores.
 
The diagram of the fusion prediction system is shown in Fig.~\ref{fig:prediction}. Specifically, we generate two top-n lists from the n-gram and TSAS models. The two lists are merged, and for each candidate content $c$ in the merged list, we normalize its n-gram and TSAS scores $S_n(c)$ and $S_T(c)$ from (\ref{eq:ngram}) and (\ref{eq:TSAS}) using max-min normalization, respectively. We then rank all the contents based on their combined normalized scores, and put the top n contents into the fused top-n list. The weight for each content is simply its combined normalized score:  
\begin{equation}
\label{eq:topn_weight}
w_c= \frac{S_n(c)+S_T(c)}{2}.
\end{equation}

\begin{figure}[htb]
    \centering
    \includegraphics[scale=0.43]{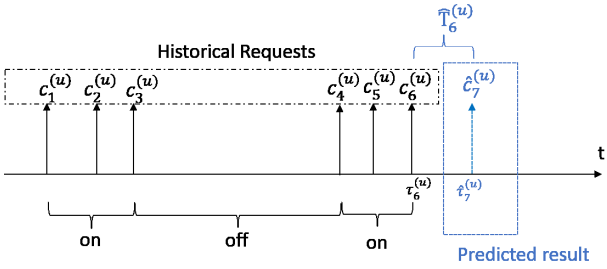}
    \caption{Predict Content and Arrival Time of Next Request for Each User with On-Off View Pattern}
    \label{fig:onoffpattern}
\end{figure}

\subsection{Next-request Arrival Time Prediction}
\label{When to Cache: Estimate Arrival Time}
For the purpose of caching, other than predicting the next content $c_{k+1}^{(u)}$, it is also important to predict when $u$ will request the next content, i.e., $\tau_{k+1}^{(u)}$. A user's activities follow  {\it on-off} pattern. When a user is actively watching videos, after finishing content $c_k$, she will generate the next content request. So the next-request arrival time is simply $\tau^{(u)}_{k+1} = \tau^{(u)}_{k} + T^{(u)}(c_k)$, where $T^{(u)}(c_k)$ is the random variable of the time duration that $u$ will watch the current content $c_k$. Meanwhile, if the user leaves the video watching session after finishing $c_k$, the next request will be generated when she becomes active again, then $\tau^{(u)}_{k+1} >  \tau^{(u)}_{k} + T^{(u)}(c_k)$, and the gap is the length of the user's off-period. The off-periods are very random, depending on lots of other factors outside of video watching. The length of an off-period can be easily hours or even days, much longer than the time-scale of edge caching. Meanwhile, a user typically watches multiple videos within each on-period, so the inter-arrivals between adjacent requests in the complete trace are dominated by the inter-arrivals between two adjacent requests within the same on-period. In this section,  we will focus on predicting the interval till the next request within the same on-period. If the next request does not arrive beyond the predicted arrival range, we will cancel our prediction and wait for the user to become active again. The next content and arrival time prediction with on-off pattern is illustrated in Fig.\ref{fig:onoffpattern}.

The key is to predict $T^{(u)}(c_k)$. One way is to use the statistics of the watching time of other users for $c_k$. However, we don't have the actual watching time in our trace. As a work-around, we use the interval till the next request after $c_k$ to approximate the watching time for $c_k$. To mitigate  the approximation error when $c_k$ is the last request of an on-period, we first cap the watching time for each video type  with a reasonable upper bound, e.g. three hours for movies. Then we use the sample median, instead of sample mean, to estimate $\mu (T(c_k))$, to limit the impact of the outliers. Similarly, we also obtain the sample variance of the watching time $\sigma^2(T(c_k))$. Then we assume that user $u$ will finish the  current content $c_k$ and generate the next content request at a uniformly random time in a future window of  $[a(\tau^{(u)}_{k+1}), b(\tau^{(u)}_{k+1})]$, 
where $a(\tau^{(u)}_{k+1}) =\tau^{(u)}_{k} + \mu T(c_k)) - \sigma(L(c_{k}))/2$, and 
$b(\tau^{(u)}_{k+1}) = \tau^{(u)}_{k} + \mu (T(c_k)) +\sigma(L(c_{k}))/2$.

It is also possible to generate more ``personalized" watching time prediction by treating $T^{(u)}(c_k)$ as $u$'s personal preference/rating for content $c_k$. We applied Matrix Factorization to estimate $T^{(u)}(c_k)$, but the estimation errors are higher than the simple sample median estimation. We will further study personalized watching time prediction in  future. 
 

\section{Caching with Per-User Content Prediction}

\label{sec:Caching}
Given the per-user next content request prediction from all users, we now aggregate them into time-sensitive predictive cache scores for hybrid proactive-reactive caching.     
\subsection{Time-sensitive Predictive Caching Score}
\label{Time-sensitive Predictive Caching Score}

\begin{figure}[htbp]
\centerline{\includegraphics[width=1\linewidth]{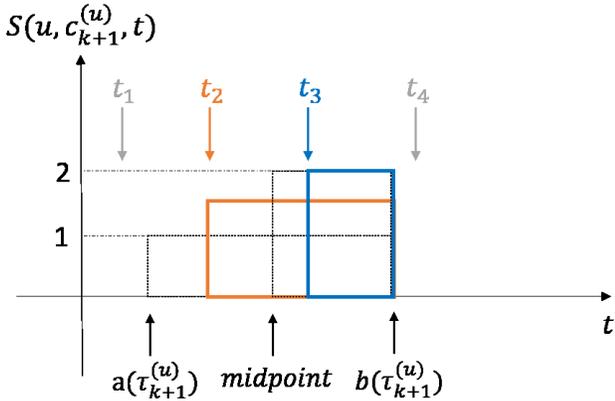}}
\caption{An example of updating predictive score as posterior probability over time. 
}
\label{fig:score}
\end{figure}

As soon as we predicted the next content $c^{(u)}_{k+1}$ and its arrival range $[a(\tau^{(u)}_{k+1}), b(\tau^{(u)}_{k+1})]$, we assign predictive caching scores to quantify its potential caching gain. This score should be time-sensitive. If the predicted arrival range is still ahead, i.e., $t < a(\tau^{(u)}_{k+1})$, it is not immediately urgent to cache the content. If $a(\tau^{(u)}_{k+1}) \le t < mid$, where $mid =  \frac {a(\tau^{(u)}_{k+1})+b(\tau^{(u)}_{k+1})} 2$. i.e., the estimated lower bound for arrival has passed, but the midpoint of the range is still ahead, it becomes  urgent to cache the content, we update our uniform prior distribution and assign a predictive caching score proportional to the posterior density function $p(\tau^{(u)}_{k+1}=t | \tau^{(u)}_{k+1} \ge t)$. 
 If the time has passed the midpoint of the estimate range, we gradually reduce our confidence about the predicted arrival. We don't update the posterior density any more, and use a fixed predictive score of $2$, which is the posterior density when $t$ reaches the midpoint, until the predicted arrival upper bound  $b(\tau^{(u)}_{k+1})$, beyond which the prediction is concluded wrong, and the predictive caching score for the predicted content is set back to zero. 
 The user $u$'s contribution to the  predictive caching score of  $c^{(u)}_{k+1}$ is updated as: 
 \begin{equation}
\label{predictivescore2}
    \mathcal S(u,c^{(u)}_{k+1}, t)=\left\{
\begin{array}{ccl}
0       &      & {t < a(\tau^{(u)}_{k+1})}\\
\frac {b(\tau^{(u)}_{k+1})- a(\tau^{(u)}_{k+1})}{b(\tau^{(u)}_{k+1})-t}      &      & {a(\tau^{(u)}_{k+1}) \le t < mid}\\
2         &      & mid \le t  \le b(\tau^{(u)}_{k+1})\\
0 & &{t>b(\tau^{(u)}_{k+1})}
\end{array} \right.
\end{equation}
An example of score update is illustrated in Fig.~\ref{fig:score}. Meanwhile, whenever $u$ requests a new content, we will generate a new prediction based on the newly requested content, the predictive caching score for the previously predicted content from $u$ will be reset, and caching score for the newly predicted content will be calculated according to (\ref{predictivescore2}).     

\begin{figure}[htbp]
\centerline{\includegraphics[width=0.9\linewidth]{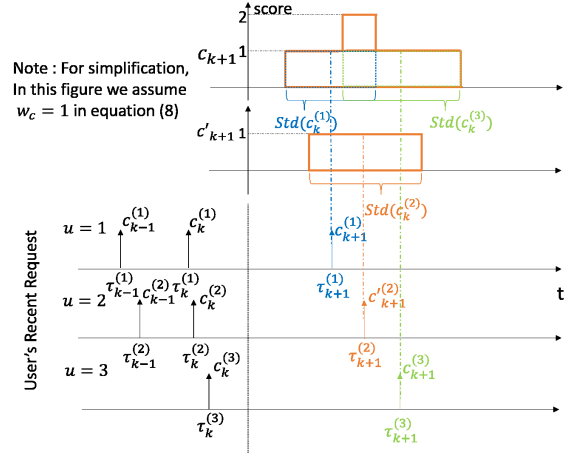}}
\caption{An example of aggregating user-content scores $S(u,c,t)$ into content predictive score $\mathcal P_1(c)$. There are 3 users and two content$(c_{k+1}$ and $c'_{k+1})$ in this example. To simplify the figure, each user only has one predicted content. User1 and user3 are predicted to request $c_{k+1}$, and user2 is predicted to request $c'_{k+1}$. $\mathcal P_1(c_{k+1})$ is the envelope of the orange distribution.}
\label{fig:conditional probabilities aggregated}
\end{figure}
 At any given time $t$, let $\mathcal A(t)$ be the set of the active users. For each $ u \in \mathcal A(t)$, based on her most recent content request sequence, we can generate the top-n list $\mathcal L_u(n,t)$ of contents that $u$ is mostly likely to request the next using the fusion model in Sec.~\ref{sec:fusion}. Each content $c$ in the top-n list has a prediction weight $w_c$ from (\ref{eq:topn_weight}). The total weighted predictive caching score for any content in the combined top-n list of all the active users can be calculated as:
\begin{equation} 
\label{eq:firstscore}
 \mathcal P_1(c, t) = \sum_{u: c \in \mathcal L_u(n,t)} w_c S(u,c,t), \quad \forall c \in \cup_{u \in \mathcal A(t)} \mathcal L_u(n,t),
\end{equation} 
 where $S(u,c,t)$ is calculated using (\ref{predictivescore2}). For TV-series data, we only predict one candidate (the next episode), i.e., n=1, and $w_c = 1$. An user-content scores aggregation example is shown in Fig.~\ref{fig:conditional probabilities aggregated}. If a content shows up in multiple users' top-n lists, but the expected arrival range has not arrived yet, according to (\ref{predictivescore2}), the content still get zero predictive score. To distinguish such a content  from a content not showing up in any user's top-n list, we give them the second class caching priority, and use its nearest predicted arrival time as the secondary predictive caching score:   
\begin{equation} 
\label{eq:secondscore} 
 \mathcal P_2(c, t) = \frac 1 { \min_{u: c \in \mathcal L_u(n,t)}  a(\tau^{(u)}_{k+1}) - t }. 
\end{equation} 
The secondary caching score decreases with the shortest time interval till the expected arrival of any active user, following the Farthest-In-Future (FIF) caching replacement policy. For comparison, $\mathcal P_1(c, t)$ has higher priority over  $\mathcal P_2(c, t)$ if $\mathcal P_1(c, t) > 0$.

\subsection{Hybrid Proactive-Reactive Caching}
\label{Reactive Makeup}

\subsubsection{Partitioning Cache into Proactive and Reactive Portions}
If the contents that are in the top-n next-content lists are not currently in the cache, one can proactively download them into the cache so that they can be directly served from the cache if they are indeed requested by a user. So the predictive caching scores are the most suited for proactive caching to achieve high hit ratio and low latency. Meanwhile, we cannot solely depend on  predictive scores to manage the whole cache. First, our  prediction models are designed for active users and active items, and the predictive scores are time-varying. As a result, at any time, only a portion of contents that are recently active have predictive scores. Sometimes they cannot even fill up the cache, such as during early morning. Secondly, our sequential prediction models work on the per-user basis, and are not designed to capture the content interest similarity cross users. For example, after user A watched $c$, the sequential model is unlikely to predict $A$ will immediately watch $c$ again. But if $c$ is indeed a popular content, other users will likely to watch it in the near future. We need to resort to the conventional reactive caching algorithms to take advantage of the homogeneity of content interests of a user group.  
\begin{table}[htb]%
\caption{Prediction Accuracy Comparison between Proactive and Reactive Caching Scores}
\label{table:top_medium_bottle}
\centerline
{
\begin{tabular}{||c||c|c|}
\hline
Recall &  Predictive score &  LRU2 \\
\hline
Top 10          &0.4750&\bf{0.5417}   \\
\hline
Top 10-50       &\bf{0.3781}&0.3448      \\ 
\hline
Top 50-100      &\bf{0.3350}&0.2321\\
\hline
Top 100-200     &\bf{0.3186}&0.1746\\
\hline
Top 200-500     &\bf{0.2548}&0.1200\\
\hline
\end{tabular}
}
\end{table}
To verify this, we conducted a case study using two days of 
user content requests in our $datasetA_1$. At the beginning of every 30 minutes, we generate a list of 2,000 contents with the highest predictive caching scores, and another list of 2,000 contents with the highest LRU-2 scores (the inverse of the time elapsed since the previous two requests for a content). Then we calculate the "recall" as the fraction of the actual contents requested within the upcoming 30 minutes that are covered by each list. Table~\ref{table:top_medium_bottle} reports the recall ratios for contents in different popularity groups. It is clear that both lists cover the more popular contents better. It is also interesting to notice that LRU2 score is better than the  predictive score when predicting the most popular contents, while predictive score is better than LRU2 for the rest of the popularity groups. The relative recall gap gets larger for the less popular groups. This suggests that our predictive scores can be used to improve caching performance for contents with medium popularity.  

We now present PEC, a hybrid caching system, that takes  advantage of the complementary prediction power of predictive and reactive caching scores. PEC partitions the cache into two portions, proactive portion and reactive portion. The proactive portion is used to prefetch contents with high predictive caching scores, while the reactive portion stores contents without predictive scores and is updated reactively using any reactive caching algorithm. For our experiments in Section.\ref{Evaluation}, we uses LRU2 for reactive caching. 


%
%
%

\subsubsection{Content Prefetching and Replacement}
\begin{algorithm}
{\fontsize{9pt}{9pt}\selectfont
\caption{Per-request Processing and Reactive Cache Replacement}
\label{alg:PEC1}
  \begin{algorithmic}[1]
 	\STATEx  {\textbf{Input:} user $u$ requests content $c_t$ at time $t$;}
	\STATEx {\textbf{Output:} updated predictive caching scores, refreshed reactive cache portion.}
        \IF {$c_t$ in cache}
				    \STATE {$hit++$}
	\ELSE
	\STATE{download $c_t$ to reactive portion, evict content with lowest reactive caching score}
	\ENDIF
        \STATE{reset user $u$'s contributions to predictive scores of contents in $u$'s last top-n list $L_u(n,t')$;}
        \STATE{add $\langle c_t, t\rangle$ to $u$'s request history, generate new top-n list $L_u(n,t)$;}
        \STATE{for each contents in the new top-n list $L_u(n,t)$, update their proactive caching scores;}      
\end{algorithmic}}
\end{algorithm}

Algorithm~\ref{alg:PEC1} describes how our hybrid caching algorithm processes each new requested content $c_t$ by user $u$ at time $t$. If $c_t$ is either in the proactive or the reactive portion, it will be directly served from the cache and counted as a new cache hit. Otherwise, $c_t$ will be downloaded from the server and stored in the reactive portion. If the reactive portion is full, the content with the lowest reactive caching score, such as the LRU2 score, will be evicted from the reactive portion. Since $u$ has just generated a new content request, the top-n next-content list $L_u(n,t')$ generated when $u$ requested the previous content at $t'$ expires. All predictive caching scores $\mathcal S(u,c, t)$ calculated for $c \in L_u(n,t')$ will be reset. We then generate a new top-n list $L_u(n,t)$ using the fusion model in Section~\ref{sec:fusion} with $\langle c_t, t \rangle$ as the most recent content request, and update the predictive caching scores of all contents in the new list according to (\ref{predictivescore2}), (\ref{eq:firstscore}) and (\ref{eq:secondscore}). 

\begin{algorithm}
{\fontsize{9pt}{9pt}\selectfont
\caption{Periodic Predictive Score Update and Proactive Cache Replacement}
\label{alg:PEC2}
  \begin{algorithmic}[1]
 	\STATEx  {\textbf{Input:} predictive scores $\mathcal P_1$ and $\mathcal P_2$ for all active contents;}
	\STATEx {\textbf{Output:} updated predictive caching scores, refreshed proactive cache portion.}
        \WHILE{true}
\IF {periodic update timer expires}
\STATE{update $\mathcal P_1$ and $\mathcal P_2$ for all active contents;}
\STATE{restart update timer;}   
\ENDIF
\IF {link to server is idle, and prefetch quota\footnotemark available}
\STATE{prefetch the content with the highest ($\mathcal P_1$, $\mathcal P_2$) score not in cache, replace the content with the lowest ($\mathcal P_1$, $\mathcal P_2$) score in proactive portion.}
	\ENDIF
 \ENDWHILE
\end{algorithmic}}
\end{algorithm}

Algorithm~\ref{alg:PEC2} describes how the predictive caching scores are updated periodically over time, and how the proactive portion is refreshed through prefetching. We periodically update the predictive caching scores of all content in the top-n lists of all active users according to (\ref{predictivescore2}), (\ref{eq:firstscore}) and (\ref{eq:secondscore}). Whenever there is a chance for prefetching, we will prefetch the content with the highest ($\mathcal P_1$, $\mathcal P_2$) scores but not in the cache 
(proactive nor reaction portion) into the proactive portion. If needed, the content with the lowest ($\mathcal P_1$, $\mathcal P_2$) score will be evicted from the proactive portion.

\subsubsection{Dynamic Partition Adjustment}
\label{dynamic partition ratio}
  In PEC, the predictive scores are time-varying, the number of contents with predictive scores can be dynamic. To avoid assigning too much storage for proactive caching when there are only a small number of contents can be prefetched, we impose the dynamic partitioning mechanism. We first set up  lower and upper bounds for proactive portion as  $\alpha*Cache\_Size$, $\beta*Cache\_Size$, with hyper-parameters $0<\alpha < \beta<1$.  At time $t$, 
  $n(t)$ is the number of contents with predictive scores. We set  $\gamma*n(t)$ as the target size for  proactive portion ($\gamma$ can be larger or smaller than one, depending on the prediction quality). After each request, $n(t)$ is updated. If the current proactive cache size is less than $\gamma*n(t)$, we increase it by one (up to $\beta*Cache\_Size$); if the current proactive cache size is larger than $\gamma*n(t)$, we decrease it by one (down to $\alpha*Cache\_Size$).

\footnotetext{Prefetch quota means the three-level bandwidth overhead controlling strategy is satisfied.}

\subsubsection{Controlling Prefetching Bandwidth Overhead}
\label{bandwid overhead control method}
{Although proactive caching gives us more freedom to update the cache and boosts the caching performance by predicting the content  popularity in the near future, it consumes extra bandwidth to prefetch content. We propose a three-level bandwidth overhead controlling strategy. 
\begin{enumerate}
    \item Firstly, prefetching is only conducted in the background. Whenever a missed content is being downloaded from the server, the prefetching is banned. In other words, prefetching only utilizes idle bandwidth to improve caching performance without interfering with the regular content downloads.  
    
    \item Secondly, after each user content request, if the link becomes idle, we allow at most one content prefetching to control prefetching traffic. 
  
    \item Thirdly, to further limit the bandwidth overhead, we can introduce {\it prefetching gap $K$} to limit the prefetching frequency. Prefetching gap is the minimum number of user content  requests between two prefetchings. For example, if $K=3$, it means a new prefetching is allowed only after three new user content requests. The bandwidth overhead and prefetching efficiency tradeoff will be studied in Section.\ref{bandwidth overhead eval}.
\end{enumerate}
}

\section{Evaluation}

\label{Evaluation}
In this section, we evaluate the performance of PEC on real world datasets from two content providers. Datasets details are introduced in Section \ref{dataset}. Prediction accuracy of per-user 
next-content and next-request-time are evaluated  in Section \ref{nextcontentprediction} and \ref{nextrequesttimeprediction} accordingly. Section \ref{cachingandpecconfigurations} presents  caching simulator setup, evaluation metrics, and PEC settings and its computation complexity. We compare PEC with several state-of-the-art reactive and proactive caching benchmarks in Section \ref{cachesimulationcomparison}. The trade-off between bandwidth overhead and proactive caching gain is investigated in Section \ref{bandwidth overhead eval}.
%
%
%
%
%

\subsection{Dataset}
\label{dataset}
The first dataset, datasetA is content request trace with timestamps. It was collected from IPTV users in different provinces of China, and each user is identified by her IP address. On average, 61.90\% of the video requests are for TV series, 24.91\% for movies,  8.94\% for TV shows, and 4.25\% for other types of videos. DatasetB was collected from users of a major OTT video service in a major city of  China. The data format is similar to datasetA, except each user has a unique ID, instead of IP address.  51.99\% of requests are for TV series and 48.01\% are for the other types of videos. datasetB  only contains active users who generates at least 10 requests each day.\\
\begin{table}[htb]%
\caption{DATASET Details}
\label{table:dataset_detail}
\centerline
{
\begin{tabular}{||c||c|c|c|c|}
\hline
Dataset &  $A_1$  &  $A_2$ & $B_1$ & $B_2$ \\
\hline
\# of users    &58,016 &5,363 &507 & 263     \\
\hline
\# of contents  &65k &20k &26k &16k  \\
\hline
\# of requests  &536k & 59k &100k & 50k \\
\hline
time span   & \multicolumn{2}{c|}{ 13 days}& \multicolumn{2}{c|}{ 7 days}\\
\hline
training set & \multicolumn{2}{c|}{ First 11 days}& \multicolumn{2}{c|}{ First 5 days}\\
\hline
testing set   & \multicolumn{4}{c|}{ Last 2 days}    \\
\hline
\end{tabular}
}
\end{table}
To emulate Edge Caching scenarios, we use a subset of datasetA based on user's IP prefix. We call it $datasetA_1$. We further sample a smaller subset, $datasetA_2$, of users sharing the same $/16$ IP prefix. Similarly, we randomly sample one subset from datasetB as $datasetB_1$, and another smaller $datasetB_2$ from $datasetB_1$. The details are shown in Table \ref{table:dataset_detail}.

\subsection{Next-Content Prediction}
\label{nextcontentprediction}

\subsubsection{Model Training and Configuration}

\begin{table}[htbp]
\caption{\label{tb:hyper_para} Hyper-parameter Configuration}
\centering
\begin{tabular}{cc} \toprule 
Parameter &  Value\\
\midrule
input sequence length for TSAS & 50  \\[-0.05em]
max time interval capping k & 5 hours  \\[-0.05em]
learning rate & 0.001  \\[-0.05em]
latent vector dimension d & 50  \\[-0.05em]
\# of self-attention blocks & 2  \\[-0.05em]
batch size & 128  \\[-0.05em]
drop rate & 0.2  \\ [-0.05em]
n-gram selection & $n=3$ \\ [-0.05em]
download time for each content & $0.5s$ \\ [-0.05em]

 \bottomrule
 \end{tabular}
\end{table}

We tried different configurations and the best one is shown in Table~\ref{tb:hyper_para}. The input of our TSAS next-content prediction model is a user's past 50 requests with timestamps. If the number of the past requests is less than 50, empty contents will be padded with the timestamp of the start of the dataset. The maximal time interval between two requests is capped at 5 hours. 
Other parameters are set as default values in Tensorflow 1.12.0. Following  practices in \cite{Tisasrec} \cite{kang2018self}, to reduce the content/user space and improve prediction accuracy, we only predict the next requests for active users and active contents, since the requests generated by active users are the most important for caching. Users having at least 3 requests and contents requested at least 3 times are considered as active. For the n-gram model, we set $n=3$ to achieve the best complexity-performance tradeoff in our experiments. 
Similarly, 3-gram model is built on active users and contents as well.\\


\subsubsection{Prediction Accuracy}
 The accuracy of the top-n next-content list is measured by the {\it top-n hit ratio}, which is defined as the fraction of predictions that the next content watched by $u$ is indeed in the predicted top-n list. For TV series, our simple heuristics only predict the next episode as the next content, i.e, $n=1$. The top-1 hit ratio is 45.31\% for $datasetA_1$ and 32.48\% for $datasetB_1$ respectively.  The prediction accuracy for non-TV videos of $datasetA_1$ are presented in Table~\ref{table:contentpredictionaccuracy_IPTV}. The upper bound for fusing top-n lists of TSAS and 3-gram is calculated as the hit ratio of the combined top-n lists of the two models. It is clear that users are less predictable when watching non-TV videos, self-attention TSAS model outperforms the 3-gram model at larger $n$. The simple CombSum fusion can significantly improve the prediction accuracy of individual models to approach the fusion upper bound. This suggests that the two prediction models are complementary, and can be easily  combined. Similarly, for $datasetB_1$,  TSAS and 3-gram also have complementary performance, and fusion hit ratios are $0.1039$, $0.1705$, and $0.2548$ for top-1, top-3 and top-10 prediction lists respectively. 
 
 
\begin{table}[htb]%
\caption{Hit Ratio for non-TV videos in $datasetA_1$}
\label{table:contentpredictionaccuracy_IPTV}
\centerline
{
\begin{tabular}{||c||c|c|c|c|}
\hline
Hit@n &  Fusion &  TSAS & 3-gram & Upper bound\\
\hline
n=10      &\bf{0.1483}    & 0.1263        & 0.1117    & 0.1603    \\
\hline
n=3       &\bf{0.1132}    & 0.0911        & 0.0944     & 0.1296  \\
\hline
n=1       &\bf{0.0779}    & 0.0691        & 0.0725     & 0.1012     \\ 
\hline
\end{tabular}
}
\end{table}


\subsection{Next-Request-Time Prediction}
\label{nextrequesttimeprediction}
As discussed in Section~\ref{When to Cache: Estimate Arrival Time}, a user consumes video following ON-OFF pattern. In each ON session, the user will request the next video as soon as she finishes the last video, and we can use the time interval between  request for $c$ and the next request in the training set to approximate the watching time for $c$. This approximation is problematic if $c$ is the last video of an ON session. To filter out such intervals, we first estimate the length of a video based on the largest mode of the distribution of the watch time approximations from all users (assuming a significant portion of users will finish watching the video). We then discard all approximations larger than the estimated video length. Finally, we take the sample mean and sample variance of the filtered watch time approximations to estimate the next request arrival range as described in Sec.~\ref{When to Cache: Estimate Arrival Time}. For a content that has never showed up in the training set, we simply use 20 mins as an estimation. The difference between the estimated mean arrival time and the actual arrival time for $datasetA_1$ is  shown in Table~\ref{table:arrivaltimeaccuracy_IPTV}. The prediction errors for $datasetB_1$ are $11.92$ minutes and $15.27$ minutes for TV-Series and non-TV-Series, respectively. Knowing that the exact arrival time cannot be very accurately predicted, the predictive score in Equation \ref{predictivescore2} is calculated using a time interval $[a(\tau^{(u)}_{k+1}), b(\tau^{(u)}_{k+1})]$ for loss tolerance. 




\begin{table}[htb]%
\caption{Next  Request  Arrival  Prediction  Error (mins) on Testing Set ( $datasetA_1$ )}
\label{table:arrivaltimeaccuracy_IPTV}
\centerline
{
\begin{tabular}{||c||c|c|c|c|}
\hline
 Type&  TV-Series &  Movie & Show \\
\hline
Error (mins)&15.5197   & 25.3991       & 12.1609   \\
\hline

\end{tabular}
}
\end{table}

\subsection{Caching and PEC Configurations}
\label{cachingandpecconfigurations}
\subsubsection{Caching Configuration and Performance Metrics}
We simulate a single edge cache with variable storage size. The edge cache is connected to a content server hosting all the contents. If the content requested by a user is in the cache, it can access it with zero latency, otherwise, the requested content will be downloaded from the server, incurring longer latency. Cache \textbf{hit ratio} is a classic performance metric. Additionally, users are directly impacted by the content retrieval latency. Since prefetching generated by proactive caching will consume some bandwidth on the link to server, it may introduce additional delays to retrieve missed contents from server. Besides, delayed hit may happen if multiple requests are requested at same time in a queue \cite{atre2020cachingdelayedhit}. So \textbf{latency} is another important metric, which indicates how long one request needs to wait till it is pushed to the user. In the experiments, we calculate latency reduction resulted from caching by comparing the latency of cache supported content retrieval with cacheless content retrieval. We simulate the content transmission on the link to the server to evaluate the total latency of serving all requests. In our simulator, content will be downloaded sequentially from the server, the active content download  occupies the whole link bandwidth. 
Following the common research practices~\cite{DRL_contentcaching,CEC_tongyu,cocktail_tongyu_ton,Prediction-Based,HybridCaching5G}, we assume each content has the same size, and the transmission time for all contents is set to  0.5s in the following experiments if no explicit declaration. 
Finally, we monitor the bandwidth utilization on the link to quantify the traffic overhead of proactive caching. 

\subsubsection{PEC Hyper-parameter Settings and Computation Complexity}
\label{sec:Partition_Ratio}
\begin{figure}[htbp]
\centerline{\includegraphics[width=0.8\linewidth]{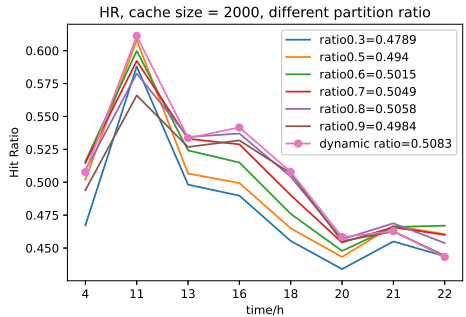}}
\caption{Hybrid Caching Performance Comparison between Dynamic Partitioning and Static Partitioning at Different Fixed Ratios.}
\label{fig:dynamic ratio}
\end{figure}

PEC employs two cache portions, with the reactive portion controlled by LRU-2, and the proactive portion updated by predictive caching scores. To maintain up-to-date  predictive scores, the periodic update timer in Algorithm \ref{alg:PEC2} is set to 5 minutes. We set the prefetching gap K as 1 in the following experiments if no explicit declaration. As discussed in Section \ref{dynamic partition ratio}, dynamic partitioning is employed in PEC. We set $\alpha=0.5,\beta=0.9,\gamma=1.2$ for $datasetA$ and $\alpha=0.3,\beta=0.6,\gamma=0.4$ for $datasetB$. 
Fig. \ref{fig:dynamic ratio} compares the cache hit ratios of dynamic partitioning with static partitioning in the first day of $datasetA_1$ testing set. Cache hit ratio is calculated for every $5K$ requests. Our dynamic  partitioning can adapt well to the user activity variations over the day and dominates static partitioning at different fixed  ratios most of the time. 

On our local computer with GTX 1660 Ti and Intel Core i5-9400, it takes 12ms to update all predictive scores every 5 minutes as described in Equation (\ref{predictivescore2}), and 23ms to make time-sensitive prediction (TSAS, 3-gram and fusion) after each user content request on average. This makes PEC implementable for real-time operation on reasonably configured edge cache nodes.\\


\subsection{Caching Experiments \& Comparison with Other Benchmarks}
\label{cachesimulationcomparison}
After training the prediction models using the training data, we now use the content requests in the testing set to conduct predictive caching experiments. 

\subsubsection{Comparison with Reactive Caching Policies} 
\begin{figure*}[t!]
 \centering{
 \begin{subfigure}[t]{0.32\linewidth}
            \centering
            \includegraphics[width=1\linewidth]{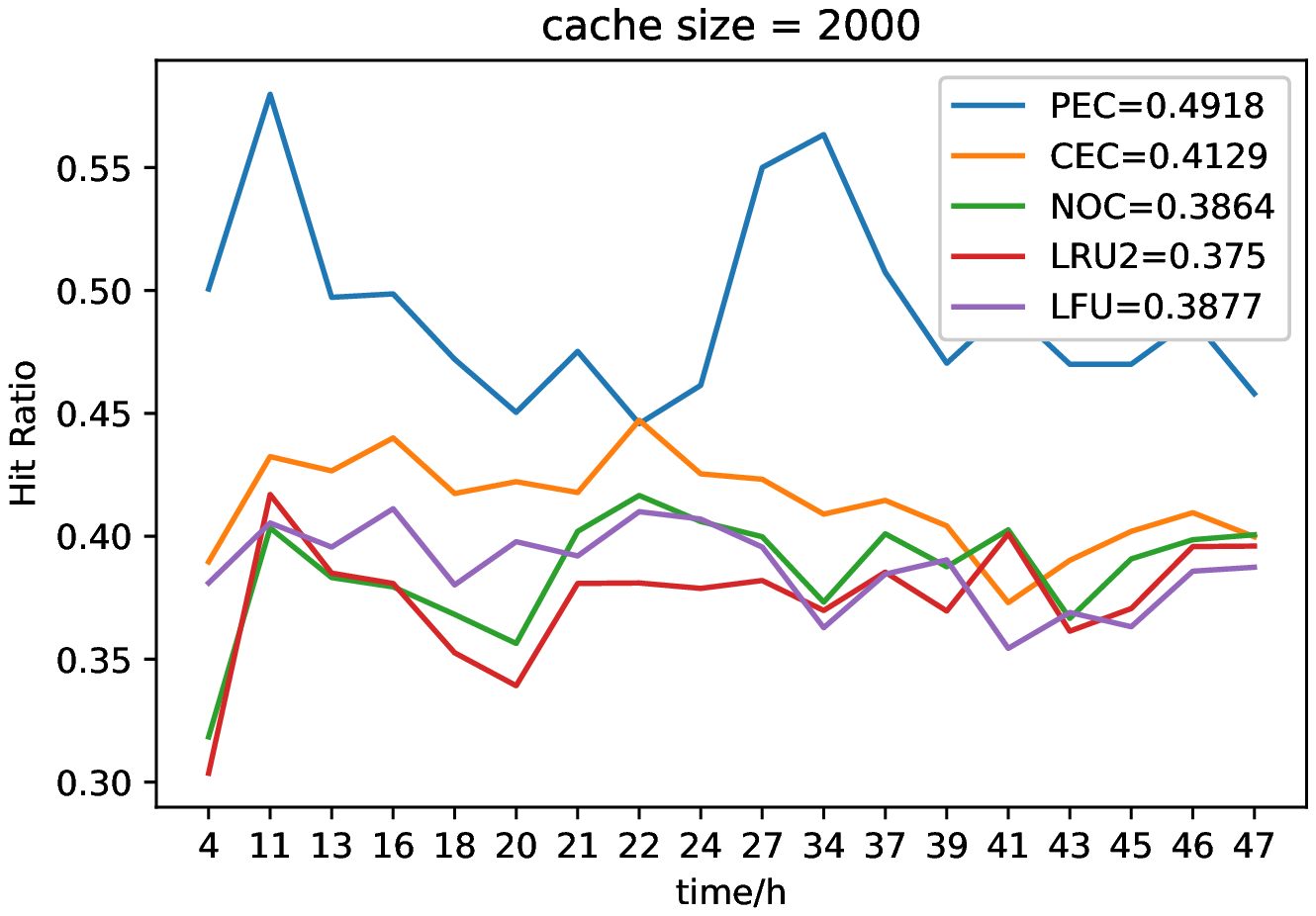}
             \caption{PEC vs. Reactive Policies}
            \label{fig:twodaycomparisonsize2k}
    \end{subfigure}%
    \begin{subfigure}[t]{0.315\linewidth}
            \centering
            \includegraphics[width=1\linewidth]{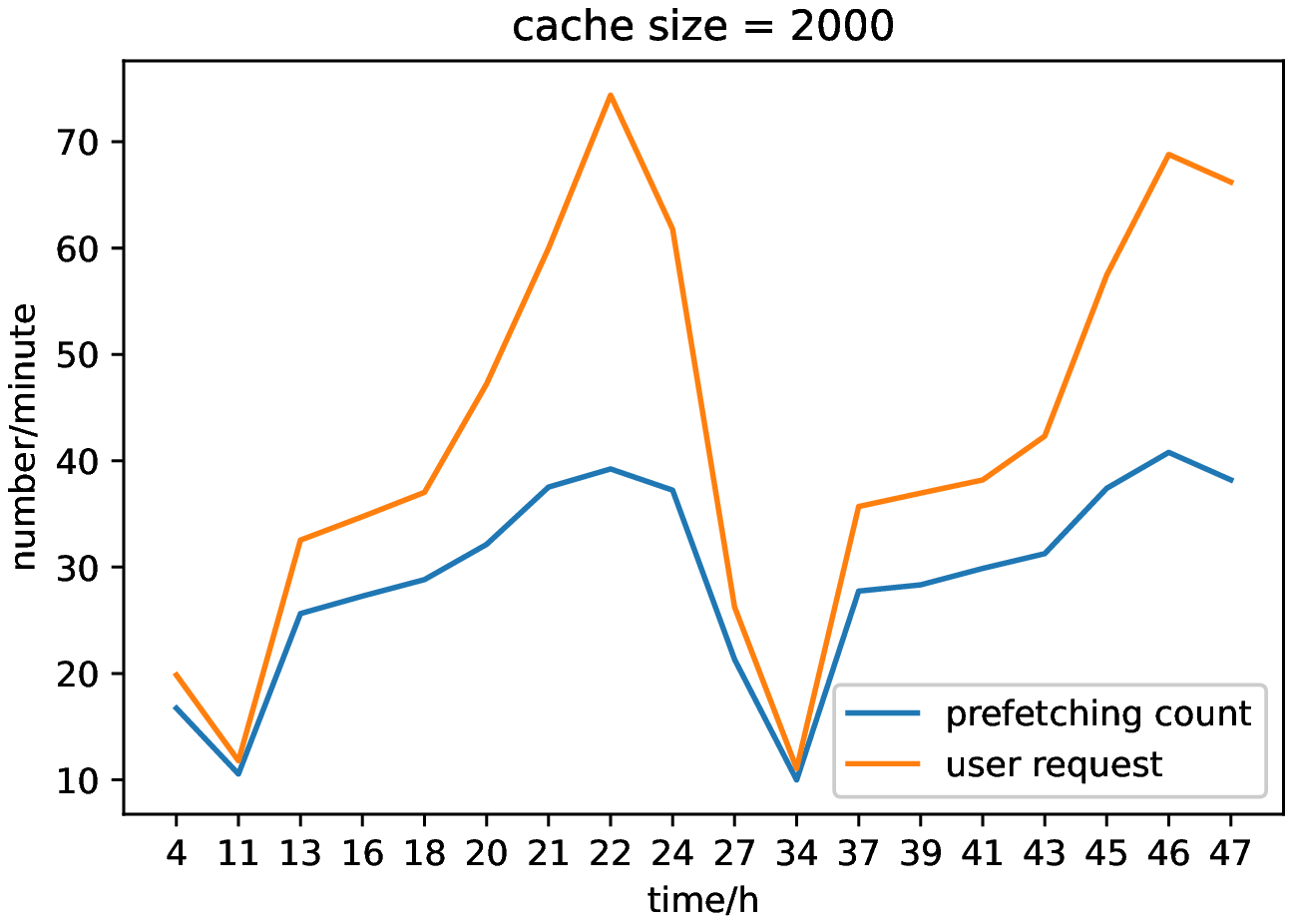}
            \caption{Number of Prefetches}
            \label{fig:prefetch time 2k size}
    \end{subfigure}
    \begin{subfigure}[t]{0.32\linewidth}
            \centering
            \includegraphics[width=1\linewidth]{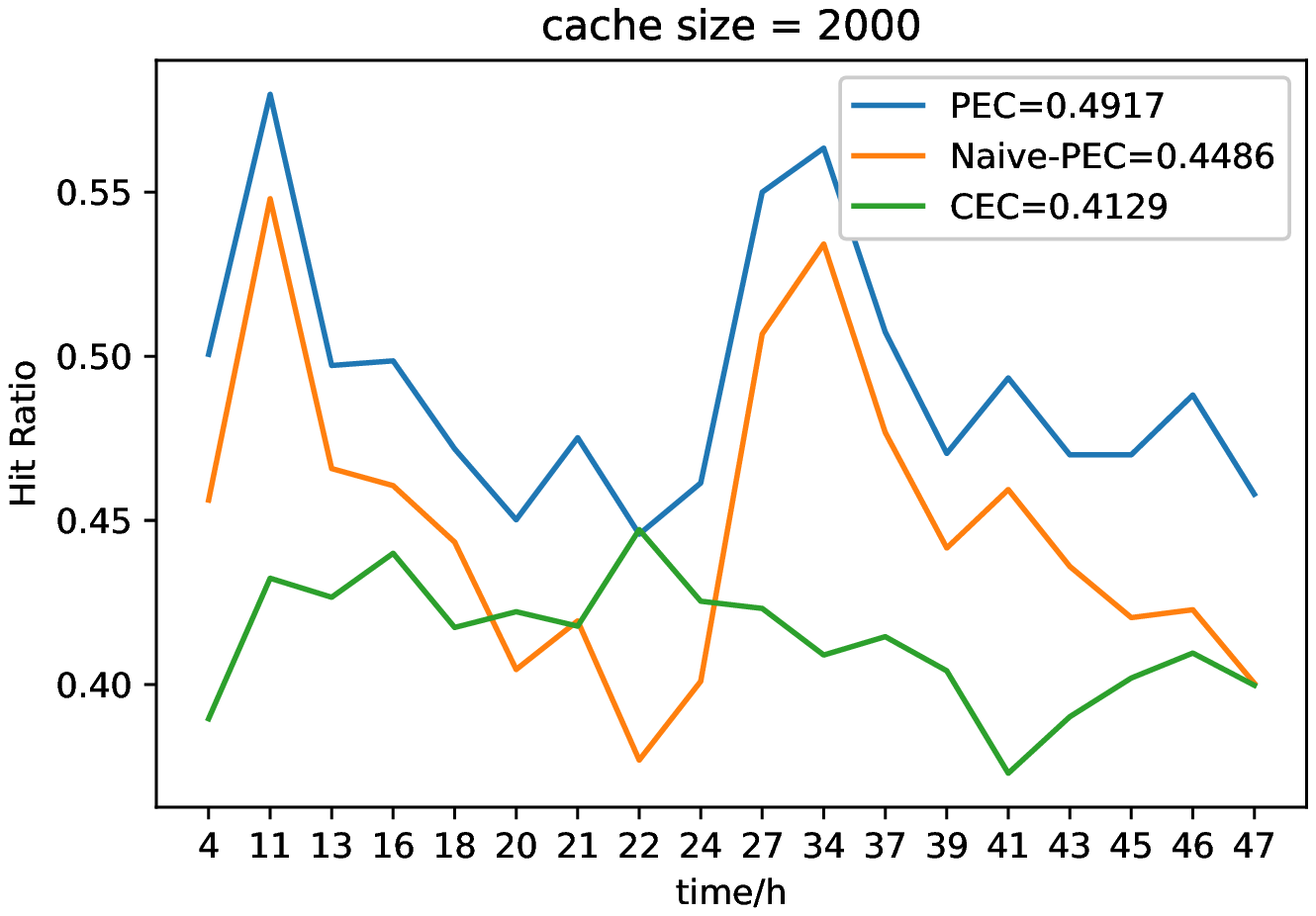}
            \caption{Fusion vs. n-gram}
            \label{fig:prediction_simple_selfattention}
    \end{subfigure}%
}
\caption{Comparison of Hit Ratio over Two Days between PEC and Reactive Caching Polices as well as Prefetching with Simple Prediction Model.}
\label{fig:Comparison with different method}
\end{figure*}

\begin{figure*}[htbp]
\centering{
\centerline{\includegraphics[width=1.3\linewidth]{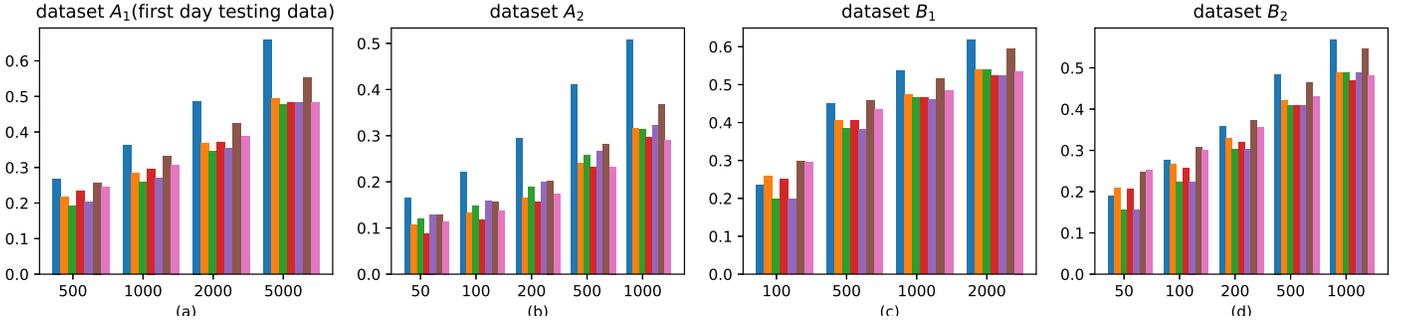}}
\caption{Hit Ratio Comparison with Benchmarks on Four Different Datasets under Different Cache Sizes}
\label{fig:hit_ratio_on_4_different_datasets}}
\end{figure*}

We first compare PEC with following reactive caching policies:
\begin{enumerate}
\item \texttt{LRU-2}: evicts content based on the time elapsed since the previous two requests. In PEC, LRU2 is also used to manage the reactive portion; 
\item \texttt{LRU}: evicts content based on the time elapsed since the last request; 
\item \texttt{LFU}: evicts content based on the request frequency in the whole history;
\item \texttt{LRB}: Learning Relaxed Belady, an online learning approach using the concept of Belady boundary~\cite{LRB_zhen}
\item \texttt{NOC}: an online learning based caching algorithm with worst-case performance guarantee~\cite{NOC_zhou};
\item \texttt{CEC}: dynamically selects reactive caching policies using reinforcement learning~\cite{CEC_tongyu};
\end{enumerate}
The results for cache size of $2,000$ over two-day testing data on $datasetA_1$ is shown in Fig.~\ref{fig:twodaycomparisonsize2k}. Cache hit ratio is calculated every 5k requests~\footnote{We couldn't customize the instantaneous hit ratio calculation of the LRB code, we only report the average hit ratio of LRB in  Figure~\ref{fig:hit_ratio_on_4_different_datasets}}. PEC has much better performance during the off-peak time, for example, from 3:00am - 15:00pm. It is because during off-peak time, there is more idle bandwidth, and PEC gets more chances to prefetch contents and update its proactive portion. Fig. \ref{fig:prefetch time 2k size} plots the user request rate and the prefetching rate by PEC, respectively. During the peak time, prefetching rate is only less than half of the request rate; during off-peak time, PEC can almost launch one prefetching after each user request, so that the proactive portion can be updated in-time to achieve high hit ratio.  
Table~\ref{table:latency datasetA_1} reports the average latency reduction of different caching polices that were adopted into our cache system simulator over the cacheless system. PEC has the largest latency reductions over all the cache sizes, thanks to its predictive prefetching. Besides, latency reduction are up to $53.23\%$, $63.48\%$, $57.85\% $ on the $datasetA_2$, $datasetB_1$, $datasetB_2$ respectively, where the latency reduction of CEC are $46.81\%$, $58.94\%$, and $52.48\%$. 
\begin{table}[htb]%
\caption{Latency reduction percentage over a cacheless system on $datasetA_1$ (first day on testing set)}
\label{table:latency datasetA_1}
\centerline
{
\begin{tabular}{|c|c|c|c|c|}
\hline
Cache size &  500 &  1000 & 2000 & 5000\\
\hhline{|=|=|=|=|=|}
PEC         &\bf{33.59\%}&\bf{44.36\%}&\bf{57.73\%}&\bf{73.79\%}   \\
\hline
LRU2        &28.71\%&37.09\%&46.51\%&60.74\%     \\ 
\hline
LRU         &27.50\%&36.48\%&45.90\%&59.28\%\\
\hline
LFU         &30.78\%&38.30\%&47.05\%&58.86\%\\
\hline
CEC         &29.78\%&43.43\%&56.02\%&67.54\%\\
\hline
\end{tabular}
}
\end{table}

We also compare PEC with these benchmarks on the other three datasets and the results are shown in  Fig.\ref{fig:hit_ratio_on_4_different_datasets}. In most cases, PEC outperforms the benchmarks, except on $datasetB_1$ and $datasetB_2$ when the cache size is small. It is because $datasetB_1$ and $datasetB_2$ are  for active users and top popular contents dominate their content requests. The traditional caching polices such as  LRU/LFU can perform well on these top popular contents. When the cache size is small, LRU/LFU can outperform PEC sometimes. When the cache size gets larger, it becomes equally important to cache medium popular contents. As demonstrated in Table~\ref{table:top_medium_bottle}, our predictive score is better than the traditional caching scores to hit contents with medium popularity. Consequently, PEC achieves much higher hit ratio on larger cache sizes. LRB focuses on large-scale datasets and considers contents with different sizes. It was not designed for edge caching. As a result, it was not sufficiently warmed up with our edge cache traces to achieve good performance.  

\subsubsection{Comparison with Simple Periodically Proactive Caching}
  The traditional proactive caching policy controls the whole storage, periodically (the period is set to 3 hours) estimate the content popularity using the past request frequencies, and load  the cache with the most popular contents in batch. There is no cache replacement between two batch updates. For a more fair comparison with PEC, 
we also implement a hybrid proactive-reactive caching policy, called modified proactive caching, which uses the same partition ratio as PEC, the reactive portion is also  controlled by LRU-2, but the proactive portion is periodically updated using past content request frequencies. The caching results are reported in Table~\ref{table:periodicalproactive} for different cache sizes on the first day of testing set of $datasetA_1$. PEC significantly outperforms the two periodic proactive cache updating policies.
The performance improvement is mostly due to: 1) predictive caching scores can better reflect the future content popularity than the statistics of the past requests, 2) PEC updates proactive scores in realtime and the proactive portion is constantly updated through background prefetching so that it can better adapt to dynamic content popularity evolution.  

\begin{table*}[htb]%
\centering
\caption{Hit Ratio and Latency Reduction Comparison with Periodic Proactive Caching}
\label{table:periodicalproactive}
{
\begin{tabular}{||C{2cm}||C{2cm}|C{2cm}||C{2cm}|C{2cm}||C{2cm}|C{2cm}||}
\hline
The First Day in Testing Set   & \multicolumn{2}{c||}{\bfseries Periodic  Proactive}  & \multicolumn{2}{c||}{\bfseries Modified Proactive}  & \multicolumn{2}{c||}{\bfseries PEC}\\
  
\hline
 {\bfseries Cache Size}   & \bfseries  latency & \bfseries HR    & \bfseries  latency & \bfseries HR    & \bfseries  latency & \bfseries HR    \\
\hline
500   &23.81\%   &0.1637  &27.48\%    &0.2020  &\bf{33.59\%}    &\bf{0.2685}     \\
\hline
1000   &29.38\% &0.2061 &34.03\%    &0.2520  &\bf{33.59\%}    &\bf{0.3624}     \\
\hline
2000   &36.65\% &0.2692 &41.81\%    &0.3212 &\bf{57.72\%}    &\bf{0.4869}     \\
\hline
5000  &47.41\%  &0.3652 &52.80\%    &0.4211 &\bf{73.79\%}    &\bf{0.6593}     \\
\hline
\end{tabular}
}
\end{table*}

\subsubsection{Comparison with Naive-PEC Driven by Simple Prediction Model.}
In PEC, we leverage time-aware TSAS model with fusion process to capture user's short-term and long-term preferences. To justify the complexity of TSAS and fusion model, we compare PEC with proactive caching guided only by the 3-gram prediction model, called Naive-PEC. To have a fair comparison, PEC and Naive-PEC use the same dynamic partition ratio, and for TV-series contents, PEC and  Naive-PEC use the same heuristic method. The only difference between them is the prediction model on the other contents. The hit ratio result is shown in Fig.\ref{fig:prediction_simple_selfattention} with the best reactive benchmark CEC as the reference line. We can notice that thanks to the prefetching method, Naive-PEC can still have a better overall performance than CEC. 
Our time-aware TSAS model with fusion process achieves  more than $10\%$ improvement over the Naive-PEC and it's worthy to use TSAS and fusion model.

\begin{figure*}[t!]
 \centering{
 \begin{subfigure}[t]{0.34\linewidth}
            \centering
            \includegraphics[width=1\linewidth]{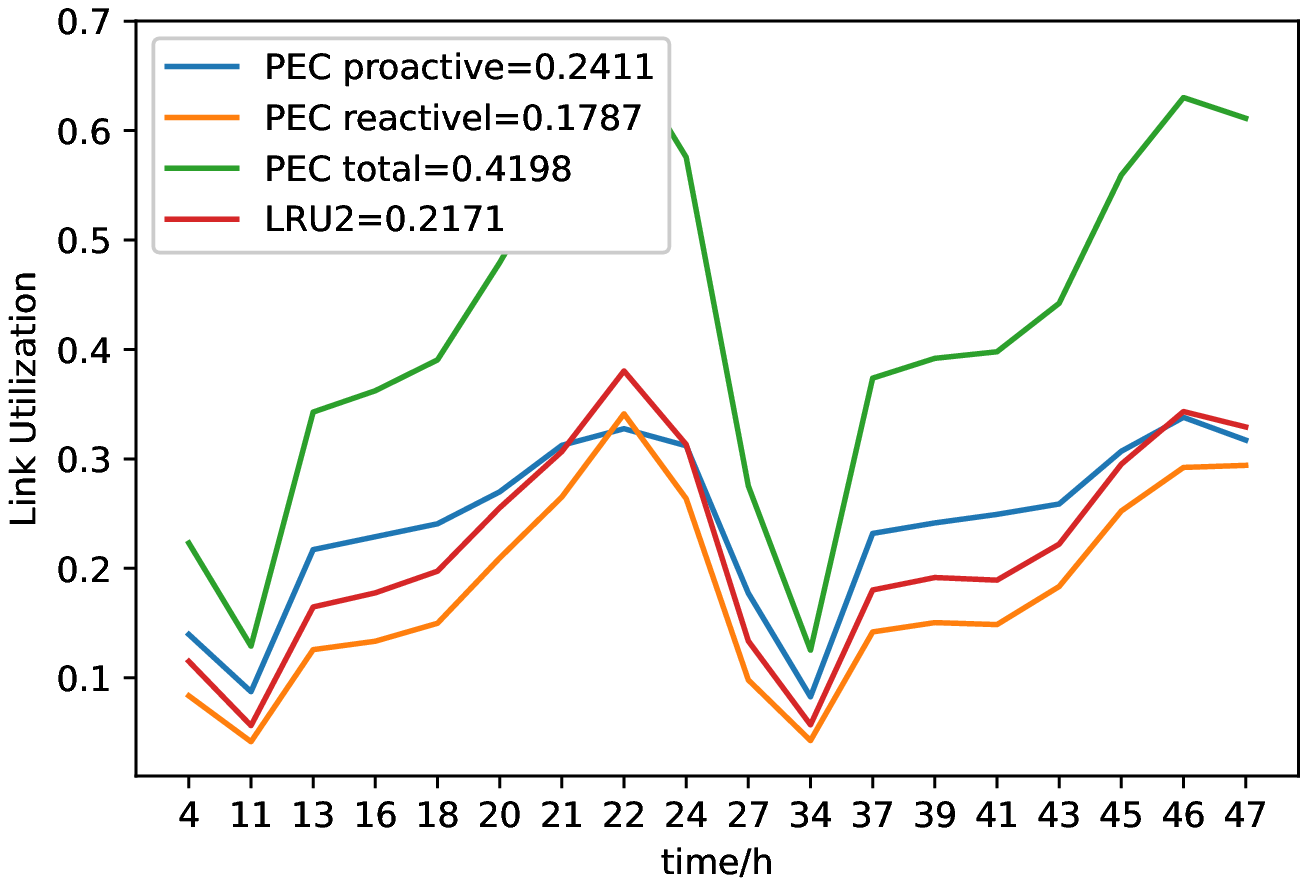}
             \caption{Nominal Link Bandwidth}
            \label{fig:linkutilization1k}
    \end{subfigure}%
    \begin{subfigure}[t]{0.34\linewidth}
            \centering
            \includegraphics[width=1\linewidth]{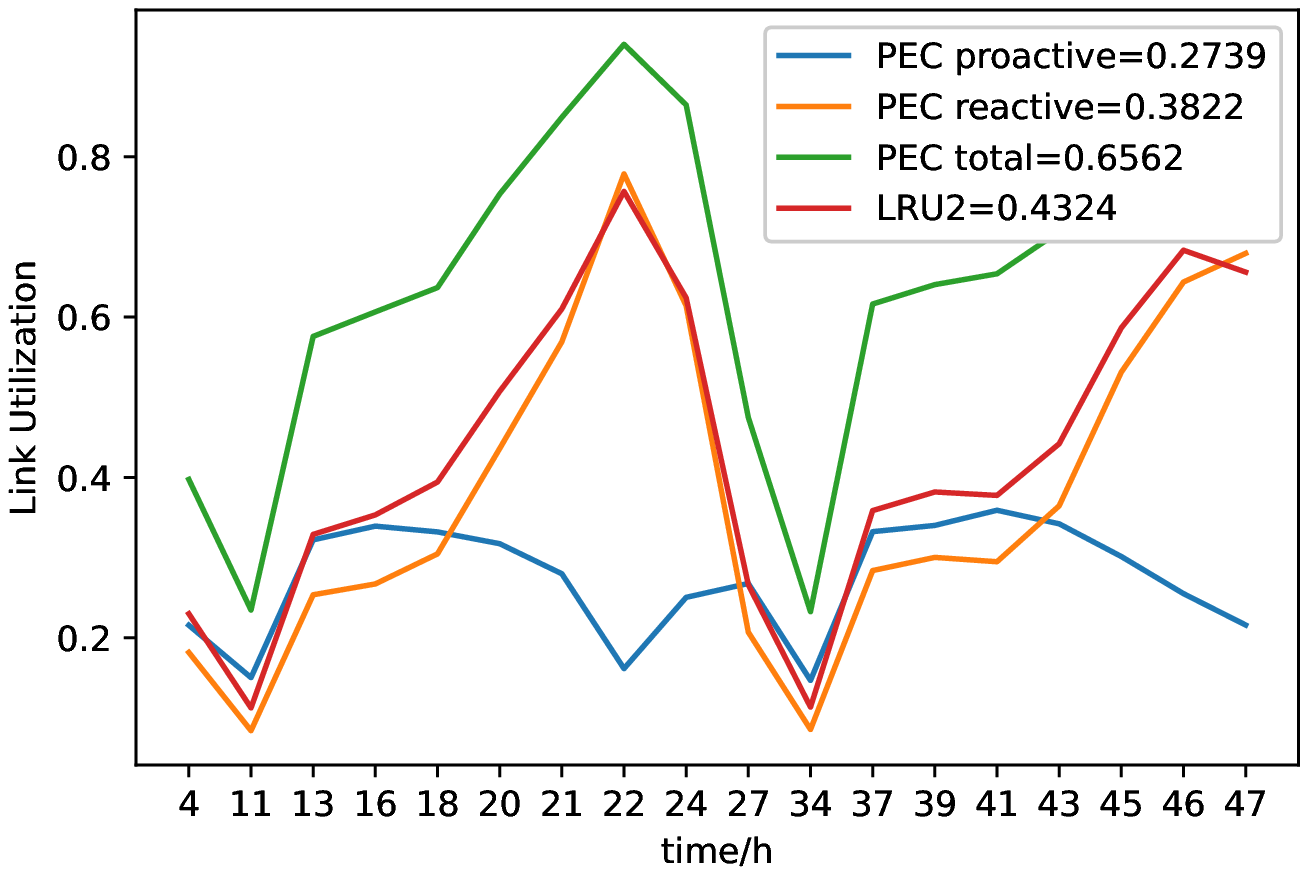}
            \caption{Halved Link Bandwidth}
            \label{fig:halfbandwidth1k}
    \end{subfigure}%
    \begin{subfigure}[t]{0.34\linewidth}
            \centering
           \includegraphics[width=1\linewidth]{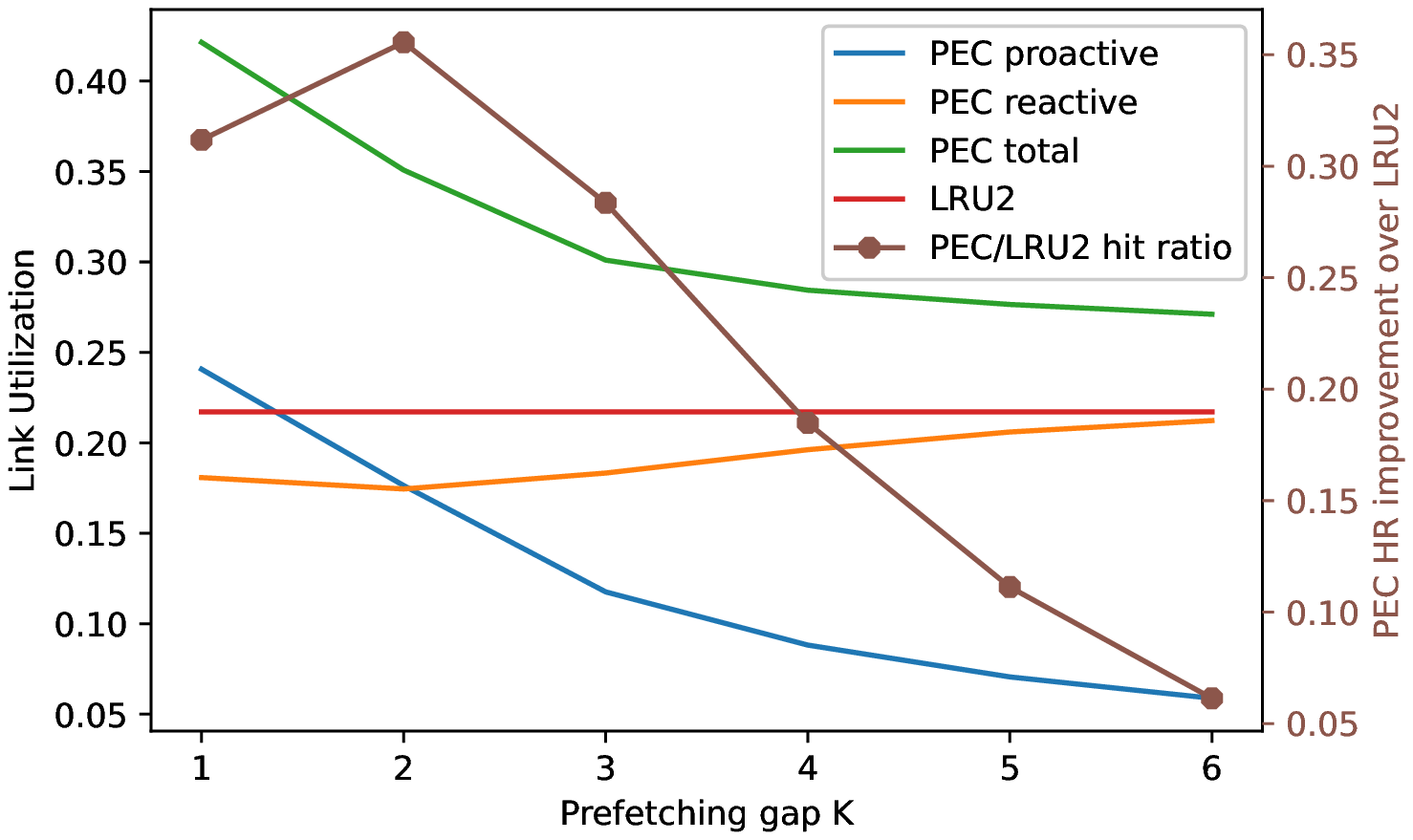}
            \caption{Impact of Prefetching Gap}
            \label{fig:prefetchingbudget_combined}
    \end{subfigure}
 }
\caption{PEC Bandwidth Overhead and Caching Gain Trade-off }
\label{fig:linkutilization}
\end{figure*}
\vspace{-0.1in}
\subsection{Bandwidth Overhead Analysis}
\label{bandwidth overhead eval}
  We report the link utilization for proactive updates and reactive updates in Fig.\ref{fig:linkutilization1k}. We also report the link utilization for the pure reactive policy LRU-2. Due to prefetching, the overall link utilization of PEC is higher than LRU-2, but is controlled within an acceptable range by the three-level bandwidth overhead control mechanism. Fig. \ref{fig:halfbandwidth1k} reports the link utilization when the link capacity is halved and then each content download takes 1 second to complete. As a result, the server spends a larger fraction of its link bandwidth to serve the missed contents, and the prefetching is suppressed to a lower fraction. This suggests that PEC can elastically adjusts its prefetching traffic to minimize its negative impact on the regular traffic. In this case, the hit ratio of PEC is $0.4632$, still higher than the LRU2 hit ratio of  $0.3718$.

 Fig.\ref{fig:prefetchingbudget_combined} shows the how the tradeoff between hit ratio and link utilization can be controlled by the   prefetching gap $K$. Larger $K$ gives less chance for prefetching, leading to lower link utilization, at the same time, degrades the performance of PEC. But PEC always outperforms LRU2 with a resealable prefetching gap. Interestingly, there is a slightly hit rate increases when the prefetching gap increases from 1 to 2. The reason can be that the predictive caching score is not perfect, if PEC prefetches into cache too many contents that never become popular, it will hurt the caching performance. But still, the difference is small, does not change the overall trend of the bandwidth-performance tradeoff. 




\vspace{-0.1in}
\section{Conclusion}

\label{Conclusion}
In this paper, we develope a novel predictive edge caching system, called PEC, which leverages on learning-based  user sequential behavior predictions and real-time background proactive content prefetches to estimate and keep track of  the highly dynamic content popularity in the near future. In our experiments driven by real-world user traces, compared with the traditional periodic proactive caching, PEC significantly improves the hit ratio by up to $80\%$, and reduces the latency by up to  $55\%$. Meanwhile, PEC also outperforms the state-of-art machine learning based reactive caching policy by $19.10\%$ in terms of hit ratio, and reduces the content retrieval latency by $9.2\%$. PEC prefetching works in the background and utilizes the spare bandwidth to boost the caching performance. Its bandwidth overhead and caching gain tradeoff can be flexibly controlled.  Our work demonstrates that per-user sequential prediction models can lead to more accurate future content popularity estimation than simple history-based statistics, and opportunistic content prefetching can be used to tradeoff spare network bandwidth for reduced latency, which is critical for the emerging edge applications. 

\printbibliography

\end{document}